# A Proposed Method for Measurement of Absolute Air Fluorescence Yield based on High Resolution Optical Emission Spectroscopy

V. Gika[1], E. Fokitis, S. Maltezos

Physics Department, National Technical University of Athens, 9, Heroon Polytechniou Zografos, ZIP 15780, Athens, Greece

**ABSTRACT**

In this work, we present a method for absolute measurement of air fluorescence yield based on high resolution optical emission spectroscopy. The absolute measurement of the air fluorescence yield is feasible using the Cherenkov light, emitted by an electron beam simultaneously with the fluorescence light, as a "standard candle". The separation of these two radiations can be accomplished exploiting the "dark" spectral regions of the emission band systems of the molecular spectrum of nitrogen. In these "dark" regions the net Cherenkov light can be recorded experimentally and be compared with the calculated one. The instrumentation for obtaining the nitrogen molecular spectra in high resolution and the noninvasive method for monitoring the rotational temperature of the emission process are also described. For the experimental evaluation of the molecular spectra analysis we used DC normal glow discharges in air performed in an appropriate spectral lamp considered as an air-fluorescence light emulator. The proposed method and the associated instrumentation could be tested and used in thin or thick target experiments in electron beam accelerators as a candidate optical system for this purpose.

## 1. INTRODUCTION

In the experimental research of Ultra High Energy Cosmic Rays (UHECRs) the air-fluorescence technique is widely used. In this case, the atmosphere essentially plays the role of a calorimeter. In particular, the fluorescence light produced by Extensive Air Showers (EAS) in the atmosphere, can be recorded by fluorescence telescopes and allows the determination of the three main parameters of the primary UHECR particles, that is, their energy, direction and composition. The Air Fluorescence Yield (AFY) is a conversion factor between the light profile and physical interpretation of the showers and expresses the number of photons produced by a charged particle per unit path length [1]. The deposited energy in the atmosphere is proportional to the energy of the primary UHECR particle. This process of energy deposition requires more detailed investigation using the experimental data and for this reason, the crucial parameter of AFY is determined in several contemporary experiments. The obtained results, in particular those from the recent measurements, are in reasonable agreement over a large electron energy range taking into account the experimental errors. However, further investigation is necessary on this issue. Recent studies have lead to somewhat more accurate measurements of the AFY but still an absolute uncertainty of nearly 12 % remains [2-5]. The AFY of the atmospheric air is much lower than that of pure $N_2$. The low yield of air is attributed to the presence of $O_2$ molecules, which due to their many low lying energy states reduce the fluorescence yield by collision de-excitation. The fluorescence light emitted by $N_2$ has its origin to the following two electronic transitions: the second-positive (2P or SPS) band system, $C^3\Pi_u - B^3\Pi_g$, and the first-negative (1N or FNS) band system of $N_2^+$, $B^2\Sigma_u^+ - X^2\Sigma_g^+$. On the other hand, each of these bands is divided into a large number of almost equally spaced spectral lines due to the presence of rotational levels. The bands usually contain the so-called "band head" due to a pile up of lines in this region belonging to P branch. The region of interest from the experimental point of view is typically from 300 and 430 nm in the near UV [5,6].

---
[1] Corresponding author
E-mail address: guika@mail.ntua.gr

The study of fluorescence light emission, by means of measurement of AFY among the emitted spectral bands, can be accomplished either using specific optical filters or using Grating Based Optical Emission Spectroscopy (OES). The optical filters must have relatively narrow spectral band-pass transmittance and thus are of interference type. Because their cut-off is shifted to lower wavelengths by the incidence angle, a very narrow transmittance curve selecting a particular spectral line is not feasible, and as a result, some neighbouring lines are also selected. On the other hand, using filters with larger pass-band, we get more optical noise in the optical band of interest, and also other undesired lines. One unexplored aspect of this topic is the study of the emitted fluorescence light using High Resolution Optical Emission Spectroscopy (HR-OES), and therefore eliminate the spectral transmittance shift of the optical filters mentioned above [7]. This technique provides higher accuracy and, at the same time, capability to monitor the rotational temperature which is the only representative thermodynamic parameter of the emission process.

The HR-OES method has been used initially by G. Davidson and R. O' Neil [8] for the analysis of nitrogen molecular spectra by using a beam of electrons of energy 5-60 keV for studying the energy-transfer processes.

The absolute measurement of the AFY using a main tool the Cherenkov light emitted by the relativistic electrons simultaneously with the fluorescence light is a favoured practice of recent relevant experiments. In the present work we describe a new method to separate the two radiations and afterwards to determine the unknown fluorescence intensity using the Cherenkov light as "standard candle". In particular, in the nitrogen molecular spectrum, beyond every band head, a non-emitting ("dark") region appears to some extent, and this feature is explained theoretically in section 4.3. In this region we can measure the Cherenkov light alone and thus the separation is feasibility. In Section 2, the definition of the air fluorescence yield parameter is described, in Section 3, we discuss the temperatures associated with the process in normal glow discharges and in particular the significance of the rotational temperature, in Section 4, the experimental evaluation of the apparatus is presented in detail, while in section 5, we describe the proposed method for absolute measurement of air fluorescence yield as it could be applied in electron beam accelerators. Finally, in Section 6 we present the conclusions.

**2. DEFINITION OF THE AIR FLUORESCENCE YIELD**

An UHECR induces in the atmosphere an Extensive Air Shower along its moving direction. The energy of the primary particle can be determined by the energy deposited in the atmosphere (calorimetric measurement). The shower is not absorbed completely in the atmosphere, and thus, the deposited energy constitutes only a fraction of the primary particle energy. The escaping energy is associated with the secondary particles reaching the ground level while there is also a small fraction of the missing energy.

From the theoretical point of view, the fluorescence yield at a wavelength $\lambda$ depends on the atmospheric pressure $p$, the temperature $T$, the humidity $u$ in the vicinity where the electromagnetic interactions are occurring, and, in principle, on the energy of the electrons or positrons. The number of fluorescence photons $dN_\gamma$ which are generated in a layer of atmosphere with thickness $dX$ registered by a fluorescence detector can be expressed as [9]:

$$\frac{dN_\gamma}{dX} = \iint \frac{dN_e(X)}{dE} \frac{dE_{dep}}{dX} Y(\lambda, p, T, u, E) \tau_{atm}(\lambda, X) \varepsilon_d(\lambda, X) dE d\lambda \quad (1)$$

where, $\tau_{atm}$ denotes the transmittance of the atmosphere, $\varepsilon_d$ the efficiency of the detector (telescope), $Y$ the air fluorescence yield (AFY), $dN_e/dE$ the energy spectrum of the electrons and the fraction $dE_{dep}/dX$ describes the energy deposited on a layer of atmosphere with thickness $dX$. In the double integral of the Eq. 1, the only variable which is a function of both the wavelength and energy is $Y$. In the case of our interest for a particular narrow energy region (the parameter $Y$ could be considered as constant), the energy terms can be taken out of the integral yielding to a simpler expression:

$$\frac{dN_\gamma}{dX} = \frac{dE_{dep}^{tot}}{dX} \int Y(\lambda, p, T, u)\tau_{atm}(\lambda, X)\varepsilon_d(\lambda, X)d\lambda \qquad (2)$$

where $dE_{dep}^{tot}/dX$ is the total (integrated) energy deposited in an atmospheric layer of thickness $dX$.

During the EAS evolution the fluorescence light is emitted in various atmospheric conditions which, in addition, vary from day to day. The model expressed by Eq. 2 is applicable in the AFY experiments of thin and thick target in electron beam accelerators, where the energy of the electrons is known.

## 3. MONITORING THE TEMPERATURE

### 3.1 The temperature in plasma

According to the theory, the laboratory plasmas usually deal with a local thermal equilibrium state instead of the thermodynamic equilibrium state. When the population of discrete energy levels follows Boltzmann's Law, it is possible to attribute for each energy distribution a corresponding temperature. The low pressure normal glow (LPNG) discharges and as well as the electron beam passing through an air chamber emitting fluorescence light are characterized as "cold plasma" (or "non-thermal plasma") sources. In this cases only a small fraction of the gas molecules are ionized, thus they are different manifestations of the local thermal equilibrium state. The gas (air) is not in equilibrium which means that the ions and neutrals are at lower temperature (normally room temperature), whereas electrons are much "hotter". Nevertheless, the rotational temperature is very close to that of the gas (kinetic) temperature. Additionally, the rotational temperature is the only reliable and remotely measured parameter expressing the dynamics of the process. Also it can be measured by a noninvasive procedure without disturbing the process itself.

For studying the effect of temperature variation on AFY, in an electron beam experiment, it is very interesting perspective to monitor the rotational temperature inside the air chamber and to investigate its correlation (linear or not) with the air temperature measured by other conventional techniques (i.e. electrostatic probes or sensors). Upon this idea we describe below the method for measuring, and thus monitoring, the rotational temperature based on HR-OES using as light source a low pressure air spectral lamp. The results can allow us to conclude about the feasibility of the method in electron beam accelerators.

**3.2 Temperatures associated with internal energy levels**

The rotational temperature of $N_2^+$ can be regarded as the temperature of $N_2$ (i.e. the plasma temperature) in good approximation under certain conditions. However, the non-thermal plasmas are typically characterized mainly by four temperatures: the electron temperature (the translational temperature of free electrons) $T_e$, the vibrational temperature $T_{vib}$, the rotational temperature $T_{rot}$ and the translational temperature $T_{trans}$. In these plasmas, created by externally applied electric fields, the above temperatures are considered to follow the inequalities: $T_e > T_{vib} > T_{rot} \approx T_{trans}$. Additionally, there is a fifth temperature, the electronic excitation temperature $T_{elex}$, which may differ from $T_e$. The modelled spectra are sensitive to $T_{elex}$, $T_{vib}$, $T_{rot}$, because the upper states of the emission transition have different energy components (electronic excitation, vibrational and rotational). The population of the upper states is given by the following equation in which it is considered that the internal energy levels follow Boltzmann distributions at $T_{elex}$, $T_{vib}$, $T_{rot}$ respectively [10,11]:

$$n_{e',\upsilon',J'} = n_{total} \frac{L}{\sigma} \frac{g_e(2J+1)\exp\left(-\frac{\varepsilon_{e'}}{T_{elex}} - \frac{\varepsilon_{\upsilon'}}{T_{vib}} - \frac{\varepsilon_{J'}}{T_{rot}}\right)}{\sum_{e,\upsilon,J} g_e(2J+1)\exp\left(-\frac{\varepsilon_e}{T_{elex}} - \frac{\varepsilon_\upsilon}{T_{vib}} - \frac{\varepsilon_J}{T_{rot}}\right)} \quad (3)$$

where, $n_{e',\upsilon',J'}$ is the population of the upper state (denoted by prime symbol) with a specific electronic excitation (*e*), vibrational (*υ*) and rotational (*J*) quantum level. The quantities $g_e$ and (2J +1) are the degeneracy (or statistical weighting) of the electronic excitation and rotational internal energy levels, respectively; the vibrational level degeneracy is equal to one. *L* is the line alteration factor due to nuclear spin and *σ* is 2 for homo-nuclear molecules and 1 for hetero-nuclear molecules. *ε* is the energy of the respective level. The summation in the denominator represents the combined rotational, vibrational and electronic excitation partition function. Each term of the form $\exp(-\varepsilon/T)$ represents the Boltzmann distribution. At atmospheric pressure, for most species, it can be assumed that $T_{rot}$ is close to $T_{trans}$ and is effectively considered as the temperature of the gas mixture because rotational to translational relaxation is fast at this pressure region [11].

## 4. EXPERIMENTAL EVALUATION OF THE APPARATUS

In order to perform HR-OES with the required accuracy for resolving the molecular band systems created by LPNG discharges, it is necessary to record the individual emission band with sufficient signal-to-noise ratio within a reasonable integration time (i.e. some minutes or one hour). Considering a synthetic model spectrum of a transition band system the rotational temperature can be determined by a multi-parametric fit applied on the rotationally resolved or alternatively on unresolved spectrum (using the Boltzmann function contour alone). The rotational temperature can be deduced as a free parameter and represents the rotational population of the light emitting species [12,13].

### 4.1 Sub-components of the setup

The prototype instrumentation we developed can be conceptually divided in four subsystems, as illustrated in Fig. 1, designed in order to serve the optimum performance in the wavelength range of interest. These subsystems are described in detail below:

*Light Source:* A spectral lamp, made by Meltz Electrolamps Moscow (MEM), with annular anode and planar cathode filled with low pressure atmospheric air (~130 mbar). It is used to provide normal glow electrical discharge, at a current of around 1 mA and under high voltage of 1450 V, emitting the characteristic molecular spectrum of nitrogen in the UV range.

*Beam forming*: The optical setup to achieve a uniform and low angular spread light beam. First, we used a UV reflecting integrating sphere (IS) which can provide a Lambertian-type emission of any light beam entering in its entrance port. Additionally, a beam collimator made by two semi-cylindrical pieces and a UV lens were used to direct and focus the light beam on the spectrograph's entrance slit.

*Spectrograph*: A stigmatic high-resolution and high-sensitivity UV spectrograph of 1 m focal length was used. It has been designed in a Czerny-Turner asymmetric configuration optimized in such a way to cancel the meridional coma satisfying the "Rosendahl condition" around a reference wavelength of 350 nm. It has relatively high f/number, of the order of 6, for reducing the astigmatism of the spherical mirrors. It uses a blaze angle grating of 1800 grooves/mm with large dimensions (100×100 mm$^2$) well adapted with that of the two spherical mirrors (each of diameter of 150 mm). Additionally, the spectrograph has been optimized in such a way to eliminate the spherical aberrations within the operating wavelength range. As a result, the wavelength resolution reached the level of 0.035 nm using an entrance slit of 30 μm. In the design of the spectrograph we have taken into account the prospect for recording the light from extended diffuse sources, like that we consider to have in the Air Fluorescence Yield experiments in electron beam accelerators. This spectrograph has been designed to record only selected narrow spectral regions corresponding to the band systems of nitrogen emission each time by moving the light detector, as it is described below.

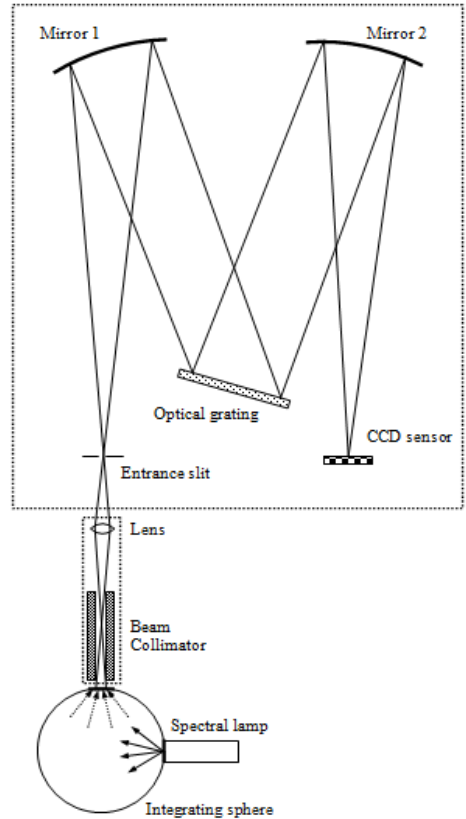

Fig. 1. The optical configuration used in front of entrance slit of the spectrograph. The light from the lamp is directed to the integrating sphere. The collimator reduces the angular spread of the beam. The lens is UV-transparent having a focal length of 250 mm.

*Light detectors*: The detectors used are two alternative CCD sensors. They also have high quantum efficiency in the near UV and are cooled at low temperatures. An appropriate mount mechanism has been designed and used for moving the CCD sensor upon a predefined optimal curve, by means of minimizing the rms of the imaging spot size. More details about the calculations and the shape of the curve are described in [14]. The first detector is a CCD sensor from SBIG[2], model ST-2000XM, consisting of 1600×1200 pixels 7.4 μm wide equipped with a thermoelectric cooling system achieving a temperature down to -15 °C. The nitrogen spectra obtained by this sensor were satisfactory, but we would like to maximize the sensitivity of the spectrograph by reducing the dark current and thus minimizing the total noise. The second detector is a CCD sensor, model CCD30-11 CCD, from e2V[3] using the AIMO technique. It consists of 1024×256 square pixels of 26 μm width with the capability to operate near $LN_2$ temperatures. Using this detector in combination with the large area of the grating and mirrors used we can accomplish the required high sensitivity of the spectrograph [7].

**4.2 Obtained molecular spectra**

Although the spectrograph can cover a spectral range of 100 nm (from 330 to 430 nm) for recording the main nitrogen molecular spectral lines, we were focused to the transition 1N(0,0) at 391.44 nm in this work. The reason is that this transition zone appears a structure of rotational lines which are easier to discriminate, and therefore, the construction of a synthetic model and the determination of the rotational temperature is feasible, as described at the next section 4.3. Our first data set concerns a spectrum obtained without the integrating sphere with e2V CCD sensor at -121 °C and 300 s integration time. This spectrum is shown in Fig. 2. The second data set concerns the spectrum obtained with the use of the integrating sphere and the CCD sensor from SBIG at -10 °C (Fig. 3). The spectrum obtained with the e2V CCD sensor appears less total noise because of the negligible read noise achieved at this low temperature. In this spectrum the standard deviation of the total noise, calculated inside the region in the absence of any signal (from 392 to 393 nm), is about 307 counts for the integration time used. The signal at the weak line at 394.3 nm is 13000 counts, with a signal-to-noise ratio equal to 43. Intensity levels similar to this line, or even lower, are expected for electron beam accelerators.

---

[2] Santa Barbara Instrument Group, 147-A Castilian Drive, Santa Barbara, CA 93117.

[3] 106 Waterhouse Lane, Chelmsford, Essex CM1 2QU, England.

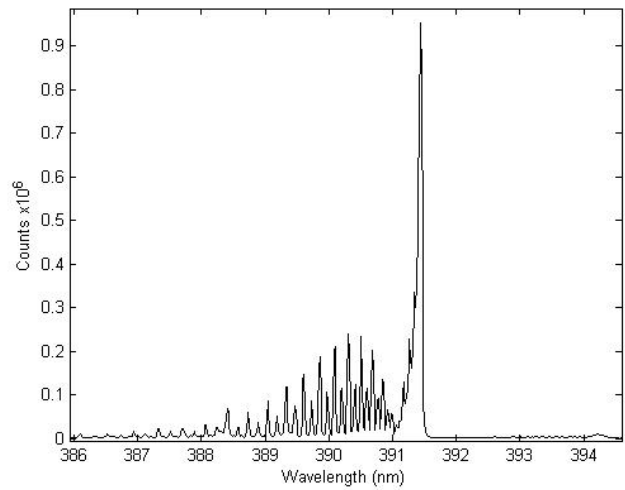

Fig. 2 Spectrum of the transition 1N(0,0) at 391.44 nm with spectral lamp alone and integration time 300 s using the CCD30-11 from e2V. The band head of the transition 2P(2,5) at 394.3 nm with its rotational structure is faintly visible while the weak band head of 1N(1,1) at 388.43 nm is also visible.

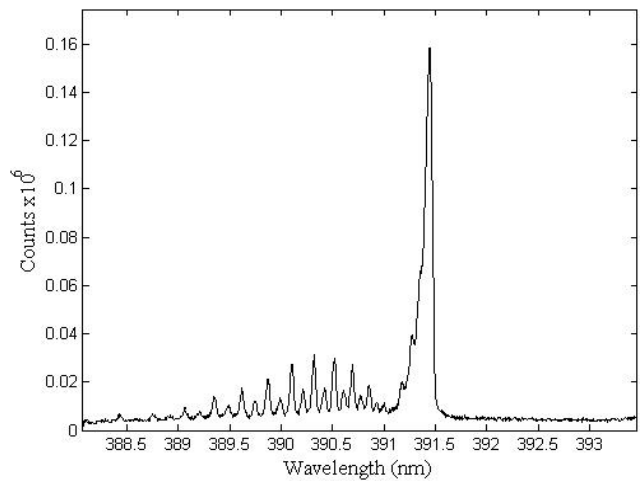

Fig. 3. The spectrum of the transition 1N(0,0) at 391.44 nm obtained using a slit of 2 mm light collimator in front of the integrating sphere and an integration time of 1 h.

On the contrary, for the band head at 391.44 nm the signal is 955000 counts leading to a signal-to-noise ratio of 3019, a value which is quite satisfactory. We note, in passing, that to our knowledge, there has not been any quantitative study in accelerator measurements of AFY concerning the weak band head of the transition 2P(2,5) at 394.3 nm which our spectrograph is capable of detecting. We have to note that, accurate relative measurements of the AFY can be accomplished by our spectrograph, by means of determining the ratio between two successive band heads, only when a pair of successive band heads (distant less than 5 nm) can be recorded. Below we give such indicative result, based on the data set obtained CCD sensor from SBIG at -10 °C, concerning the band heads at 380.5 and 375.6 nm which results to a ratio 1.51±0.05 (syst.) ±0.01 (stat.). The systematic error is minimized due to the simultaneously recording. It is important to compare this result with that given by the AIRFLY experiment using a medium resolution commercial spectrograph using the air fluorescence emitted by the electron beam in an air chamber, which is 1.52±0.08 (syst.+stat.) with 1.5 MeV electrons at 600 Torr air chamber with temperature 293 K [15]. The consistency of the results allows us to conclude about the invariance of the air fluorescence spectrum using two different excitation processes (air glow discharge and electron beam interaction respectively).

**4.3 Data analysis method**

For analyzing the experimentally obtained spectrum of the molecular nitrogen of the transition 1N(0,0) at 391.44 nm with the CCD sensor from SBIG at -10 °C, we applied the method of non-linear multi-parametric chi-square fit. The theoretical model we used is based on the structure of the lines produced by the selection rules for the rotational quantum number leading to the so-called R and P branches. From the fit we can determine the following free parameters: the rotational temperature, the resolution of the spectrograph (FWHM) and, also, the parameters used for adapting the intensities and the precise positions of the spectral lines. In order to avoid local minima in the fit, it is crucial to estimate their position with a precision of the order of the spectral resolution of the spectrograph. The theoretical expression describing the coupled rotational-vibrational transition follows a Boltzmann distribution. The recorded peaks are affected by the instrumental broadening described by a gaussian function. The general formula of the intensity for R or P branch, respectively, as a function of the rotational quantum levels can be written as follows [16]:

$$I = C \sum_{J'=0}^{N} g(J')(2J'+1) \exp\left[-\frac{hcB_{v,in}J'(J'+1)}{k_B T_r}\right] \exp\left[-\frac{4\ln 2 [\lambda - \lambda(J') + \Delta\lambda(J')]^2}{(\delta\lambda)^2}\right] \quad (4)$$

where $B_{v,in} = BJ'(J'+1) - DJ'^2(J'+1)^2$ is the rotational constant expression at the vibrational level $v$ including non-harmonic behavior with $D$ to be the centrifugal distortion constant (equal to $5.92 \times 10^{-6}$ cm$^{-1}$ in the case of nitrogen molecules), $J'$ is the rotational quantum number for R or P branch respectively, $g(J')$ is the statistical weight (equal to 1 for odd $J'$ and 1/2 for even $J'$, $\lambda(J')$ is the peak wavelength at the individual quantum number, $\Delta\lambda(J')$ is a hypothetical uncertainty of the peak wavelength, $\delta\lambda$ is the instrumental spectral resolution, $T_r$ is the rotational temperature, $k_B$ is the Boltzmann constant, $C$ is an arbitrary constant and $N$ is the number of the recorded peaks. The peak wavelengths are determined roughly by a peak finding algorithm and are introduced in the fitting model. By this method the minimization procedure uses a starting point inside the valley of the under investigation minimum, thus avoiding trapping into local minima. The hypothetical uncertainties $\Delta\lambda(J')$ play the role of correction terms in the preselected wavelength values. A more simple analysis method, based on a fitting exactly on the peaks of odd values of $J'_R$ (envelop defined by the Boltzmann function) could be alternatively used in the cases of spectra with few experimental points per peak (i.e. when the pixel size of the detector is relatively large). Nevertheless, this simple method involves larger uncertainties in the rotational temperature determination.

The above Eq. 4 is valid for R branch and as well for P branch. The total intensity should be the summation of the two terms: $I_{R,P} = I_R + I_P$. Therefore, we have to specify the sequence of the R and P quantum numbers, $J'_R$ and $J'_P$ respectively. This can be deduced from the so-called Fortrat diagram [17] shown in Fig. 4, where the lower level rotational constant $B_v$ and upper rotational constant $B_{v+1}$ in the case of nitrogen molecules have the values 2.083 cm$^{-1}$ and 1.932 cm$^{-1}$ respectively according to literature [18]. The band head is located is obtained from the condition of maximizing the function:

$$\Delta v = f(m) = (B_v + B_{v+1})m + (B_v - B_{v+1})m^2 \qquad (5)$$

where $m = J'_R + 1$ for R branch and $m = -J'_P$ for P branch, represent a pseudo quantum variable, frequently used in the Fortrat diagram. The maximizing value ($m_H$) is obtained by the formula:

$$m_H = -\frac{B_v - B_{v+1}}{2(B_v + B_{v+1})} \qquad . \qquad (6)$$

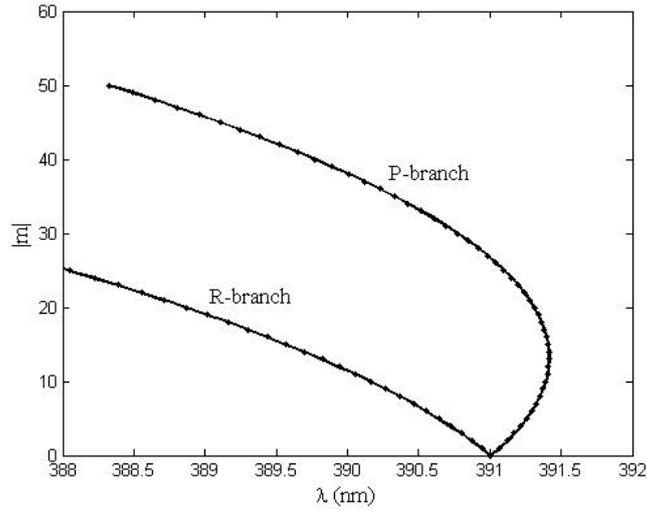

Fig. 4. The "fortrat" diagram for the band head at 391.44 nm of the FNS band system of $N_2^+$. The band head corresponds to $|m|=13$ and is extended from $J_P=1$ (were $|m|=0$) to practically $J_P=17$.

Using the values referred to above we find $m_H=-13$, leading to a band head location corresponding to $J'_P$ =13. However, the structure around the band head is developed in the range from $J'_P=1$ to $J'_P=13$ and further from $J'_P=14$ up to practically $J'_P=17$ along the symmetrical branch. The next lines (for $J'_P>17$) are visible in the obtained spectrum and essentially constitute the folded-back portion of the rotational P branch. This portion of P branch overlaps with the blue-degraded R branches (because $B_v > B_{v+1}$ which start to appear only for wavelength less than 391 nm (where $m=0$), as it is obvious from the missing peak.

This overlapping is a characteristic property of the transition of $N_2^+$ FNS and allows easier determination of the rotational temperature. The relation between R and P branches can be expressed as $J_P = J_R + 27$. For $J_R = 0$ we have $J_P = 27$ and thus the R and P branches are now clearer. The fitting curve is required to have a similar form as that of Eq. 4 in which we have specified the set of the free parameters used. For each peak we need 2 free parameters (intensity factor and the wavelength location uncertainty) while the instrumental spectral resolution is an additional free parameter, common for all lines. We applied the above fitting procedure on 31 rotational lines including the band head and using 64 parameters in total.

The first step of the fitting procedure is to identify the peak locations by an appropriate algorithm with an accuracy of 1/10 of the resolution of the nominal spectrograph. In our case, this accuracy is of the order of 0.004 nm. The fit result for the spectrum of the 391.44 nm line is shown in Fig. 5.

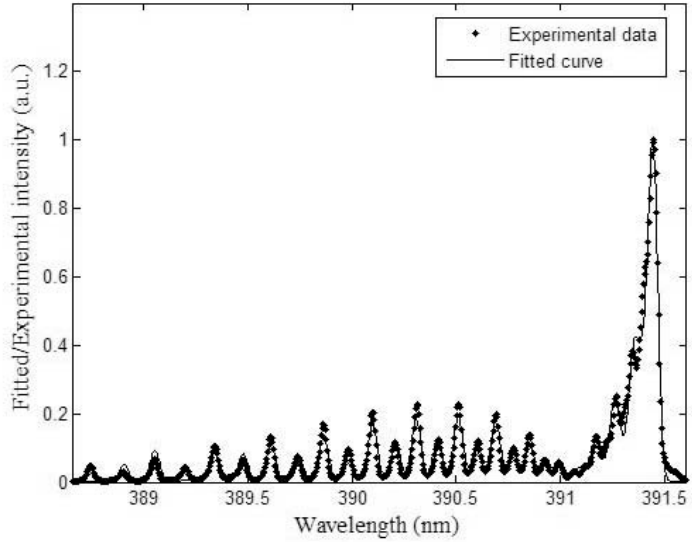

Fig. 5. Fitting of the 1N(0,0) transition at 391.44 taking into account all the 31 rotational lines, included as well as the band head.

**4.4 Discussion of the results**

The results obtained from the fitting concern mainly the free parameter of the rotational temperature. From the fit of the whole rotational spectrum (including the band head of 391.44 nm), we found $T_r = 330$ K. The uncertainty of $T_r$ is estimated of the order of 5 %. The main advantage of our method for determining $T_r$ is the use of the complete rotational spectrum (R and P branches) in the fitting model. Another important feature is the very precise representation of the structure of the experimental spectrum, due to the narrow pixel size, including about 20 points per peak. It is useful to compare the rotational temperature obtained by different plasma sources and electron beam producing air fluorescence found in the literature (the number of points per peak varied from about 4-8). The only way to demonstrate such a comparison is to apply the simple method of fitting exactly on the peaks of odd values of $J'_R$ based on Eq. 4 without including the gaussian function and using only two parameters. By this method we analyzed the spectrum of the same 1N(0,0) transition at 391.44 nm obtained by several authors (see Table I) acquiring the data from their plots by a graphical way. The obtained temperature values based on Refs [19, 20] and present work are in good agreement within the errors while that based on Ref. [8] differs by about 32 K. This discrepancy can be explained by the fact of using electron beam instead of plasma source. Therefore, the more accurate method we used for the determination of the rotational temperature can help to clarify the levels of this temperature at the electron beam accelerators. Because of the high resolution of these spectra, the exposure time in an electron beam accelerator setup to measure the AFY by our apparatus has to be of the order of few hours.

| Reference | Fluorescence light source | Spectral resolution $\delta\lambda$ (nm) | Result $T_r$ (K) |
| --- | --- | --- | --- |
| G. Davidson et al. [8] | Electron beam (energy 5-60 keV) | 0.012 | 412.8 ± 21 |
| A. Chelouah et al. [19] | Cold plasma discharge | 0.030 | 434.5 ± 32 |
| S. Maltezos et al. [20] | LPNG discharge lamp | 0.010 | 444.8 ± 17 |
| Present work | LPNG discharge lamp | 0.035 | 452.4 ± 20 |

Table I. The rotational temperature calculated by the same fitting method on molecular nitrogen spectra recorded by different sources and instrumentation.

## 5. THE METHOD FOR ABSOLUTE MEASUREMENTS

An absolute measurement of the AFY in electron beam accelerators can be accomplished obtaining the spectra of the transition band systems of interest and comparing them with a reference light source. According to [3], the Cherenkov light can be used as a reference light source ("standard candle") due to the possibility to calculate its intensity theoretically. Our proposed method is based in the use of the integrating sphere (IS) for collecting the emitted fluorescence light as it has been also used for the setup described at Ref. [21]. The IS has to be installed inside an air chamber in electron beam accelerator where the fluorescence light can be introduced via an input port and, after being trapped for several reflections on IS walls, it will be directed to an exit port. The Cherenkov light will also be diffused inside the IS and superimposed to the fluorescence. In another input port, we can install the spectral lamp for testing the setup, and as well as, to calibrate the intensity of the lamp used. In this configuration, shown in Fig. 6, the light collection efficiency from both sources is expected to be identical due to the operation principle of IS used and because their spectral structure matches quite well. The angular distribution of the fluorescence light and the lamp needn't to be taken into account in the calculations because of their intensity integration inside the IS.

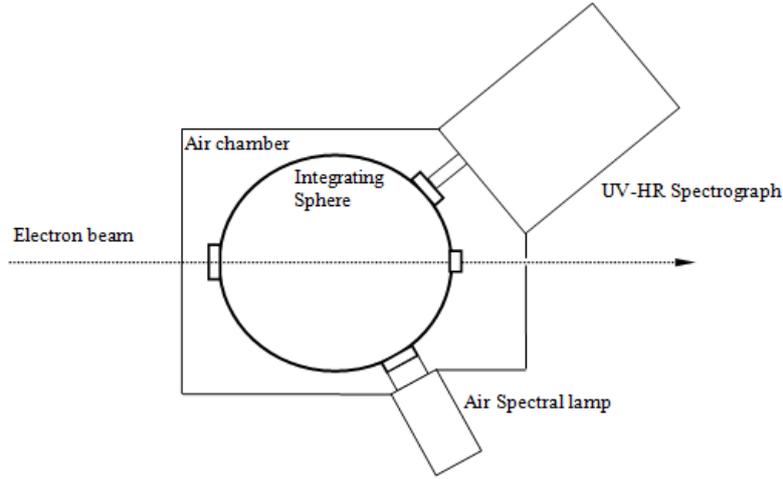

Fig. 6 A diagram illustrating the air chamber with the integrating sphere installed inside.

Referring to the Fig. 3 in section 4.2, where the spectral line of 1N(0,0) at 391.44 nm is illustrated, we consider two individual spectral regions: the region included the fluorescence spectrum, $w_F = \lambda_0 - \lambda_1$ (with $\lambda_0 = 391.65$ nm and $\lambda_1 = 389.25$ nm) and the region included the "dark" where only the net Cherenkov light is recorded, $w_C = \lambda_2 - \lambda_0$ (with $\lambda_2 = 394.00$ nm). The wavelengths $\lambda_1$, $\lambda_0$ specify the limits of the spectral line of 1N(0,0) emission spectrum and the wavelengths $\lambda_0$, $\lambda_2$ specify the limits of the "dark" region. During recording of the fluorescence light within the spectral window $w_F$, the Cherenkov light is recorded simultaneously. Taking into account the fact that in each recorded fluorescence line there is superposition with Cherenkov light as we discussed above, below we explain the methodology to spectrally disentangle the two radiations.

Let us consider the following quantities associated with the spectrum considered:

$N_{T_{obs}}$ : total number of observed photons, including fluorescence and Cherenkov photons, within the spectral window $w_F$.

$N_{C_{obs}}$ : the (net) observed number of Cherenkov photons within the spectral window $w_C$.

$N_{C_{th}}^{w_C}$ : the theoretically expected number of Cherenkov photons within the spectral window $w_C$

$N_{C_{th}}^{w_F}$ : the theoretically expected number of Cherenkov photons within the spectral window $w_F$.

$N_{F_{th}}$ : the expected number of emitted fluorescence photons within the spectral window $w_F$, which is the unknown quantity to be calibrated.

The ratio of the total signal observed in the $w_F$ region to the observed signal in the $w_C$ (net Cherenkov) region is expected to be equal to the ratio of the expected signal in the $w_F$ region (fluorescence plus Cherenkov) divided by the expected Cherenkov signal in the $w_C$ region. It is crucial that the overall efficiency of the optical system (integrating sphere, spectrograph with the associated optical components) and any other systematic uncertainty associated with the detector should be identical for both measurements because of the same spectral window of the measurements. Therefore:

$$\frac{N_{T_{obs}}}{N_{C_{obs}}} = \frac{N_{F_{th}} + N_{C_{th}}^{W_F}}{N_{C_{th}}^{W_C}}. \tag{7}$$

In this relationship the expected Cherenkov signal can be calculated as seen below.

Solving for the unknown $N_{F_{th}}$ we obtain:

$$N_{F_{th}} = \frac{N_{T_{obs}}}{N_{C_{obs}}} N_{C_{th}}^{W_C} - N_{C_{th}}^{W_F}. \tag{8}$$

According to theory of Cherenkov emission, caused by a charged particle (Ze) in atmospheric air given by Frank and Tamm, the number of emitted photons per unit path length $x$ within the spectral region from $\lambda_1$ to $\lambda_2$ is:

$$\frac{dN_{ph}}{dx} = 2\pi\alpha Z^2 \int_{\lambda_2}^{\lambda_1} \left(1 - \frac{1}{\beta^2 n^2(\lambda)}\right) \frac{d\lambda}{\lambda^2} \tag{9}$$

where $\alpha = e^2/\hbar c \approx 1/137$ is the fine structure constant, $n(\lambda)$ the refractive index of air, which can be considered constant for the spectral width of the band head considered, at this spectral region and $\beta$ the particle velocity $\left(\beta = \frac{v}{c} = \sqrt{1-\left(\frac{E_o}{E}\right)^2}\right)$. The above result is valid if the condition of excess of the particle velocity threshold, $\beta n(\lambda) > 1$, holds. The total (integrated) number of emitting photons can be obtained easily assuming that the index of refraction $n$ is essentially unchanged in this narrow spectral range in hand as follows:

$$N_C = 2\pi\alpha Z^2 \sin^2\theta_c \left(\frac{1}{\lambda_1} - \frac{1}{\lambda_2}\right) = 2\pi\alpha Z^2 \left(1 - \frac{1}{\beta^2 n^2}\right)\left(\frac{1}{\lambda_1} - \frac{1}{\lambda_2}\right). \tag{10}$$

The quantity $E$ is the particle total energy and $E_o = mc^2$ is its rest energy. According to Eq. (10), the Eq. (8) can be written as:

$$N_{F_{th}} = 2\pi\alpha Z^2 \left(1 - \frac{1}{\beta^2 n^2}\right)\left[\left(\frac{1}{\lambda_o} - \frac{1}{\lambda_2}\right)\frac{N_{T_{obs}}}{N_{C_{obs}}} - \left(\frac{1}{\lambda_1} - \frac{1}{\lambda_o}\right)\right] \quad (11)$$

where $\Delta\lambda_C = \lambda_o - \lambda_1$, $\Delta\lambda_F = \lambda_2 - \lambda_o$.

Assuming, reasonably, that $\Delta\lambda_C \ll \lambda_o$ and $\Delta\lambda_F \ll \lambda_o$ and the electron as a charge particle ($Z=1$) we can obtain an expression, which is a good approximation of $N_{F_{th}}$:

$$N_{F_{th}} \cong \frac{2\pi\alpha}{\lambda_0^2}\left(1 - \frac{1}{\beta^2 n^2}\right)\left[\Delta\lambda_C \frac{N_{T_{obs}}}{N_{C_{obs}}} - \Delta\lambda_F\right] \quad (12)$$

or by taking into account the relationship $\beta = \sqrt{1 - \left(\frac{E_o}{E}\right)^2}$ we have:

$$N_{F_{th}} \cong \frac{2\pi\alpha}{\lambda_0^2} \cdot \frac{[(n^2-1)/n^2]E^2 - E_o^2}{E^2 - E_o^2}\left(\Delta\lambda_C \frac{N_{T_{obs}}}{N_{C_{obs}}} - \Delta\lambda_F\right). \quad (13)$$

From the above equation one can also calculate the energy threshold condition for Cherenkov radiation in air as follows:

$[(n^2-1)/n^2]E^2 > E_o^2 \Rightarrow E_{thr} = nE_o/\sqrt{(n+1)(n-1)} \approx E_o/\sqrt{2(n-1)}$, where we used the approximation $n \approx 1$. It is obvious that the threshold is a decreasing function of the refractivity ($n-1$). For electrons in normal temperature and pressure conditions the energy threshold is around 21 MeV. In order to reduce the threshold we could use different gas with greater refractivity (i.e. fluorocarbons). In this case the Cherenkov light should be used for end-to-end calibration of the spectrograph while the fluorescence light will be measured in a subsequent stage with air in the chamber.

Below we give an indicative calculation of $N_{F_{exp}}$ using typical values for the spectral regions which might be used. Let, $\Delta\lambda_C = 2$ nm and $\Delta\lambda_F = 2.5$ nm in the vicinity of the wavelength $\lambda_o = 391.7$ nm. Using also the known constants, $E_o = 0.511$ MeV, $(n^2-1)/n^2 = 1.000566$ ($n = 1.0002832$ at STP), $\alpha = 1/137$, we obtain:

$$N_{F_{th}} \cong 597.84 \frac{0.000566E^2 - 0.2611}{E^2 - 0.2611}\left(\frac{N_{T_{obs}}}{N_{C_{obs}}} - 1.25\right) \text{ with } E \text{ in MeV}. \quad (14)$$

For $E^2 \gg E_o^2$ (i.e. $E$ of the order of few hundred MeV or greater), and the fraction containing the energies tends to the constant value 0.000566, and thus, the Eq. (14) can be approximated as follows:

$$N_{F_{th}} \cong 0.3384 \left( \frac{N_{T_{obs}}}{N_{C_{obs}}} - 1.25 \right). \tag{15}$$

Before measuring the two quantities $N_{T_{obs}}$ and $N_{C_{obs}}$ appearing in Eq. 15 a spectral calibration test of the apparatus has to be preceded. This test is necessary in order to accurately specify the regions $w_F$ and $w_C$ in the CCD sensor pixel matrix and can be accomplished by using the air spectral lamp. Let us now make the hypothesis of installing the setup. The uncertainties of $N_{T_{obs}}$ and $N_{C_{obs}}$ count rates comes mainly from statistical fluctuations $\delta N_{T_{obs}}$ and $\delta N_{C_{obs}}$, respectively, summed in quadrature to find the uncertainty of the ratio $N_{T_{obs}} / N_{C_{obs}}$ which is:

$$\delta \left( N_{T_{obs}} / N_{C_{obs}} \right) = \sqrt{\left( \delta N_{T_{obs}} / N_{T_{obs}} \right)^2 + \left( \delta N_{C_{obs}} / N_{C_{obs}} \right)^2} = \sqrt{1/N_{T_{obs}} + 1/N_{C_{obs}}}. \tag{16}$$

Therefore, the uncertainty of the determined fluorescence rate will be $\delta N_{F_{th}} \cong 0.3384 \delta \left( N_{T_{obs}} / N_{C_{obs}} \right)$. Any systematic effect relating to the efficiency of our apparatus is identical for both quantities, and it is cancelled in the above ratio. Below we give first a preliminary estimate of the statistical uncertainty $\delta N_{C_{obs}} = \sqrt{N_{C_{obs}}}$. Let us now to make the hypothesis for installing a IS with an indicative diameter of 150 mm (similar with that we used) in the electron beam of the AIRFLY experiment [4]. The photon rate of Cherenkov light emitted inside the IS by using the beam performance of the accelerator, that is, its intensity, $10^{10}$ e/bunch, and its repetition rate, 50 Hz, has been calculated and found $2.53 \times 10^{10}$ ph/s. Our proposed apparatus has an overall detection efficiency $(1.0 \pm 0.1) \times 10^{-8}$ counts per incident photon, and therefore, the recorded rate in the CCD sensor will be equal to $N_{C_{obs}} = 2.53 \times 10^{10} \times 1.0 \times 10^{-8} = 256$ counts/s with a relative uncertainty of 10 %. Thus, its relative uncertainty $\delta N_{C_{obs}}$ will be 0.5×10 %=5 %, resulting an absolute value 13 counts/s. Assuming zero dark noise and a negligible read noise of the $LN_2$ cooled CCD sensor: for a measurement lasting 5 min, the relative statistical error for $N_{C_{obs}}$ is 0.4 % and for 1 h is 0.1 %. Also, the relative statistical error for $N_{F_{th}}$ is 0.5 % and for 1 h is 0.15 % (assuming the photon rate of fluorescence light to be the same as the Cherenkov, and thus, a factor $\sqrt{2}$ was used in this case). These absolute measurements can be performed alternatively using a position sensitive photomultiplier located at the two pre-defined spectral regions.

Because these regions are well defined, it is not necessary to resolve the detailed structure but only to integrate the light signals. This technique could allow shorter exposure times and at the same time higher signal-to-noise ratios.

## 6. CONCLUSIONS

In this work we introduce and propose a method for absolute measurement of the air fluorescence yield based on high resolution optical emission spectroscopy. The associated instrumentation based on a high sensitivity and stigmatic UV spectrograph dedicated to this purpose can be used in electron beam accelerators to obtain the nitrogen molecular spectrum in selected spectral regions. Analysing these spectra we can determine the rotational temperature of the process, which is the most representative temperature. In order to evaluate the method and to prove the feasibility of the rotational temperature monitoring we used a spectral lamp as a air fluorescence emulator concentrating to the spectrum of band head at 391.44 nm. The proposed method for absolute measurement of AFY in electron beam accelerators based on Cherenkov light is demonstrated. We also argue that the simultaneously measuring the fluorescence and Cherenkov light, used as "standard candle", is feasible promising relatively small systematic uncertainties.

**ACKNOWLEDGMENTS**

We have to thank our Ph.D students, P. Fetfatzis and N. Maragos and as well the Msc Physicist G. Koutelieris for their contribution for the procedure for using the CCD sensor in $LN_2$ temperatures. We kindly thank Professors E. Liarokapis, I. Raptis of NTUA and their team for supporting us with their high vacuum facility. We also thank Mrs. K. Holland and Professor A. Holland from XCAM for their general support.